\documentclass[12pt]{article}
\usepackage{graphicx}
\usepackage{cite}
\usepackage{amssymb,amsmath}
\begin{document}
\title{Conformational impact on deformation of DNA TATA-box}%
\author{P.P. Kanevska, S.N. Volkov\\
Bogolyubov Institute for Theoretical Physics, NAS of
Ukraine,\\14-b Metrolohichna Str., Kiev 03143, Ukraine \\
snvolkov@bitp.kiev.ua }\maketitle
\setcounter{page}{1}%
\maketitle
\begin{abstract}
The theoretical study of deformability of  special sequence of DNA
double helix TATA-box is presentated. The paper elaborates on the
mechanisms of abnormal deformation of DNA TATA-box double helix
that cannot be explained using the standard mechanical model of
polymer molecules (WLC) and needs more detailed modeling.
Analyzing of DNA TATA-box deformation it is shown the molucule can
undergo significant deformations due to its property of the
structural polymorphism, that is, possibility of the double helix
fragment to exist in more then one conformations. In addition to
elastic components (bending, twisting), the presented model
includes the following deformation features: possibility of
conformation rearrangement of the shapes of the sugar rings,
effects of a specific nucleotide sequence and anisotropy, the
coupling between components. Presented model allows describe
abnormal deformation based on physical special  fitures of  double
helix inner structure.
\end{abstract}

\section{Introduction}
\label{intro}

The key properties of DNA functioning are contained in the
structure of specific nucleotide sequences.  Regulatory gene
fragments have specific sequence pairs.  One of these fragments is
TATA-box, has a constant part of TATA and plays a major role in
the transfer of genetic information, determining the beginning of
reading the information recorded in the gene. TATA-box-binding
with a certain part of the regulatory protein complex
(TATA-binding protein, TBP) initiates the formation of a complex
that signals the place of the beginning of reading genetic
information and is responsible for the accuracy of protein 1
synthesis \cite{KimBurley94}.
\begin{figure*}
\includegraphics[height=45mm]{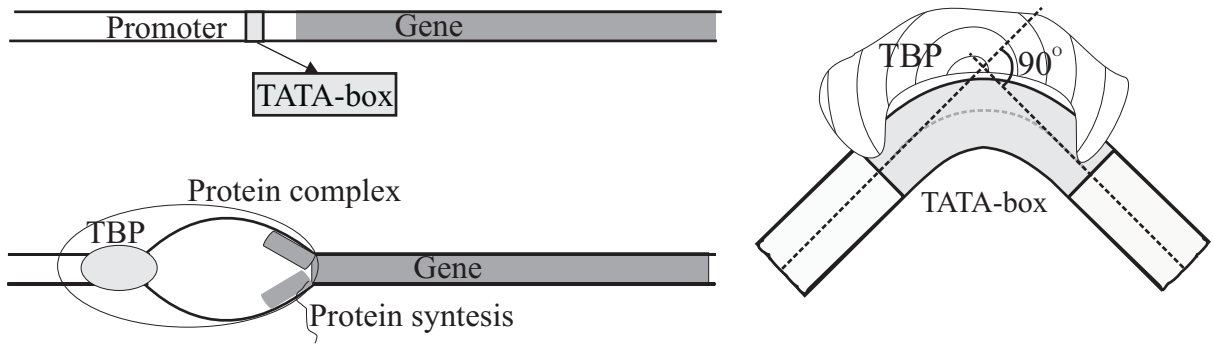}
\caption{Schematic representation of the mechanism for reading
genetic information. Deformation of the TATA box upon binding to
TBP leads to the formation of a DNA-protein complex, the formation
of which determines the beginning of information reading from DNA
during protein synthesis}
\end{figure*}

Figure 1 shows a schematic representation of the mechanism for
reading genetic information, which begins with the recognition and
deformation of the DNA TATA box.  The structure of the TATA box
associated with TBP was solved with high accuracy more than 20
years ago for the first time \cite{TATA93_1,TATA93_2}.

In the TATA box, significant deformations were observed without
destruction of the double helix.  The total deformation of the
central part was approximately $90^o$ bend in the direction of
major groove of the double helix and a $90^o$ unwinding. So there
is a direct interaction of the regulatory protein with the
informational part of the double helix, the base pairs of the
fragment.

At the same time, the alternating purine-purimidine sequence poly
(AT) does not show specific bendability in the free state. The
main feature of the fragment is bistability of the conformation of
the sugar ring found experimentally \cite{Klug_79,Brahms_81} even
without binding to other molecules.

In the case of deformation of the TATA box bound with TBP, the
property of conformational bistability is most clearly manifested
due to certain boundary conditions. The formation of the complex
of the TATA-box with TBP is characterized by intercalation of
parts of the TBP molecule at the ends of the DNA fragment, while
different forms of the sugar ring are observed in pairs adjacent
to the intercalator.

According to the WLC model, which is usually used to describe DNA
deformation, the deformation energy of the TATA box is greater
than the energy of its melting and should have destroyed the
double helix.  Based on the fact that the TATA box retains the
structure of the double helix in the complex with TBP, remembering
that the formation of this complex determines the base pair from
which transcription begins, it can be assumed that the TATA box
must have specific physical characteristics that provide
significant localized deformation with a recognition accuracy of
up to 1 bp. \cite{Klug93}.
\begin{center}
\begin{figure*}
\includegraphics[height=48mm]{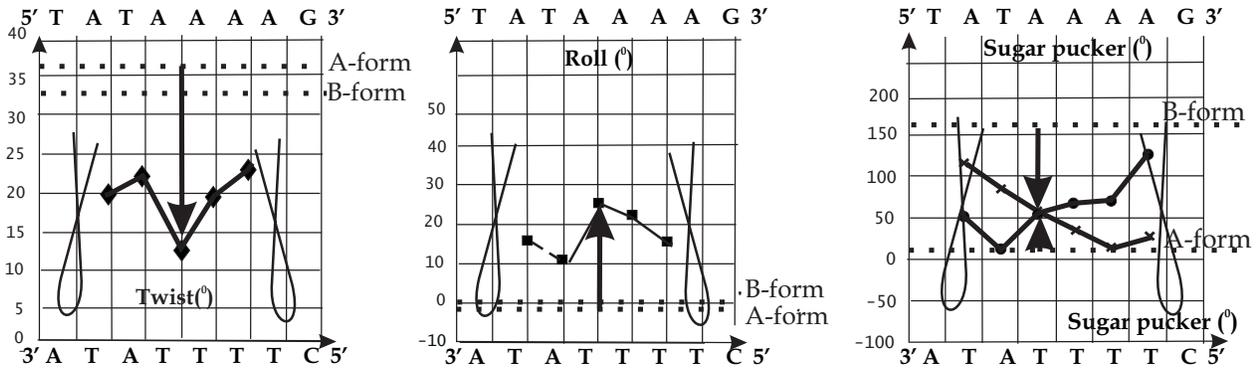}
\caption{The values of the dominant parameters of the central
fragment TATA boxing. a)Twist of each base pair step in deformed
TATA-box; b)Roll of each base pair step in deformed TATA-box;
c)Sugar ring pucker of each nucleotide in deformed TATA-box. The
arrows show the dominant deviations in the center of the TATA-side
for the dominant parameters} \label{twist}
\end{figure*}
\end{center}

In order to clarify the physical nature of high bendability of DND
TATA-box, short fragments of the double helix of different
sequences were investigated; however, it was not confirmed very
special elastic stiffness of any sequences \cite{Vologodskii}. The
difference in the values of the stiffness constant for different
sequences is no more than 10\%. The softening of such a value is
not enough to provide bend like TATA-box undergoes. Greater
flexibility of the double helix and, accordingly, a decrease in
stiffness can be provided by local destruction of the double helix
under stress\cite{Volgodskii_13}. Experimental data on the
formation of cycles with a length of 158 bp. did not show the
localization of deformation on the TATA sequence \cite{Amzallag}.

At the same time, the process of deformation of the TATA box upon
interaction with TBP is described as sequential intercalation with
subsequent approximating of the double helix bend to the shape of
the saddle protein \cite{Burley}. The physical mechanism of the
formation of significant deformation localized in the sequence of
the TATA-box DNA remains unclear.

From our point of view, a significant TATA-specific change in the
double helix conformation can occur in this fragment, which leads
to an effective change in rigidity. Thus, due to its physical
properties, a special DNA sequence exhibits a physical level of
information, which is recognized by TBP in a sequence capable of
achieving the deformation necessary and improbable for other
fragments so that the process of starting the reading of genetic
information begins \cite{Klug93}. Nevertheless, despite almost 30
years ago this conclusion, the completed physical interpretation
has not been formulated.

We analyzed the changes in the structure of the double helix of
the TATA box of DNA with certain boundary conditions, which are
realized in a complex with the TBP protein, the structural parts
of which are embedded between base pairs at the ends of the
fragment.  The most characteristic changes in the structure of the
double helix are emphasized as dominant parameters: torsion,
bending and change in the shape of the sugar ring.  It was found
that significant deformations correlate with sugar ring form
change. It should be noted that such interrelation has remained
unattended so far has not been specially studied.

However,  NMR studies showed that the limitation of the mobility
of sugar by replacing it with an inflexible analogue is reflected
in the DNA flexibility\cite{Bax}. Molecular modeling confirmed the
correlation of structural parameters with a gradual change in one
or both forms of sugar \cite{Nesterova}.

In order to build relevant deformation model of the TATA box, it
is necessary to take into account the conformational degree of
freedom reflecting the mobility of the sugar ring in DNA and its
correlation with another deformation parameters.  In this work, a
model of deformation that takes into account the conformational
mobility and its correlation with bending was constructed, an
abnormally large localized deformation of DNA was obtained
associated with conformational rearrangement of the form.

Obtained deformation does not require the destruction of the
double helix, hydrogen bonds and stacking, as is characteristic of
the deformed TATA-box of DNA upon interaction with TBP.  It has
been shown that the deformation is advantageous when the bending
siffness of the fragment decreases by about 10\%, which is
probably realized due to the neutralization of the charges of the
fragment through interaction with the protein complex
\cite{Lavery02}  and under specific boundary conditions of the
conformational component, which arise due to the penetration of
protein loops intercalated at the ends of the TATA box.

\section{Modeling of deformation induced with sugar ring form
changing  of the TATA-box of DNA}
\subsection{Structural changes in the TATA-box of DNA in a complex
with TBP}

In order to describe the physical level of information set in the
specific deformability of a DNA fragment, we  analyze the
structure of the TATA-box of DNA in a complex with TBP.  The
regulatory fragment TATA-box consists of 8 base pairs and
undergoes significant structural changes during recognition and
protein binding.  The interaction of the TATA box with TBP is
accompanied by inserting (intercalation) of the structural
elements of protein (phenyl-alanine loops) between 1st and 2d, as
well as 7th and 8th base pairs of the DNA fragment.  Intercalation
leads to a change in the structural parameters of the DNA base
pair step adjacent to the intercalator.

Moreover, the same changes are cooperatively taking place in
neighbor base pairs inside the TATA-box.  The greatest changes
occur in the following parameters of the double helix structure
(dominant parameters of TATA-box deformation): the slope between
adjacent pairs, roll increases, the twist angle decreases, and the
sugar shape of $5'$- nucleotides of base pairs, which are
contacting with the intercalator changes. The amplitude of these
changes dramatically exceeds the amplitudes of thermal
fluctuations in the selected parameters.  On the figure 2 one can
see significant deviations in the dominant parameters from the
values of the double helix forms characteristic for A and B. Loops
schematically depict intercalators, and for neighbor bp steps,
changes in the twist and roll parameters and the sugar ring of the
fragment nucleotides are noticeable; however, in the middle of the
fragment, the maximum deviation from the values in the A and B
forms is observed.

Simultaneously observing the value of the pseudorotation angle for
each nucleotide of the double helix, one can see changes in each
strand of the double helix within the central fragment from close
to $C2'-endo$ to close to $C3'-endo$ in the $5'-3'$ direction in
both strands. The figures show the values of the parameters
presented in the work \cite{KimBurley94}. For other sequences of
the TATA box, the type of deformation is the same, but the
quantitative values may differ for different sequences.
\cite{TATA93_1,TATA93_2,Klug93,Dickerson96}.

As a result, the TATA box manifests itself as a sequence that
contains a certain conformational mobility, which manifests itself
under fixed boundary conditions and induces deformation of the
fragment. Deformation  associated with a change in conformation of
sugar ring pucker is called intrinsically induced deformation
\cite{Crothers}. Obviously, the special mobility of the structure
of the double helix of the TATA box cannot be described within the
framework of a linear harmonic approach in modeling. Significant
changes indicate a nonlinear deformation mechanism. At the same
time, nonlinearity is contained in the bimodality of the shape of
the sugar ring potential function of the TA-sequence; therefore,
the inclusion of a conformational component describing sugar
changes may explain the formation of a static bell-like
deformation of the TATA box.

\subsection{Dominant components of the deformation model
TATA-boxing DNA}

DNA double helix is polymorphic macromolecule and which exhibits a
number of structures that arise  during its functioning under
external action equal to forces of piconewton order
\cite{Bustamante2003}. In particular, TA DNA sequences are known
to exhibit the coexistence of both $C2'-endo$ and $C3'-endo$ forms
of sugar as fixed by NMR \cite{Klug_79} and Raman
\cite{Brahms_81}. Analysis of various DNA sequences also confirms
the bimodality of TA DNA sequences in a number of parameters of
the \cite{Orozco12} double helix structure. Deviations from the
canonical forms were also observed in the molecular dynamics of
the TATA box \cite{Lavery98}.

Thus, the AT-rich TATA box is a conformationally bistable complex
in which the shape of sugar rings in the chains of the double
helix is different and the dominant change can be a simultaneous
change in the shape of sugars in both chains. Moreover,
intercalated TBP loops at the ends of the TATA box fix different
sugar conformations at the ends of the fragment. Study of the
effect of modifying intercalators using molecular dynamics showed
that the modification affects the amount of bending at the
intercalation site, but not in the central part of the
fragment\cite{Sticht}.

Thus, the deformation of the central part of the TATA box is
determined rather not by elastic deformation at the ends, but by
changes in the conformations of sugars in the nucleotides adjacent
to the intercalators. The conformational potential of the
pseudorotation angle of the sugar ring of the DNA nucleotide has
two minima of different depths, which determines the double-well
shape of the potential energy of the TATA box pair. The minima of
the conformational energy of the pair correspond to states of
pairs at the ends of the fragment, fixed by intercalators.

For a theoretical description of DNA deformation under conditions
when conformational changes in the double helix occur, let us
consider DNA as a chain of links with external and internal
degrees of freedom. The external degrees of freedom describe the
deformation of a molecule as a homogeneous chain and are modeled
by an elastic rod with 3 degrees of freedom: bend and twist -
bending and torsional displacements of neighboring links,
stretching - change the distance between adjacent links.

Displacement of structural elements within links described by an
internal or conformational degree of freedom. Each link is a pair
of nucleotides in which the change in the conformation of the
sugar ring occurs along a certain conformational trajectory of the
mutual arrangement of atoms (pseudorotation angle) and has two
minima of potential energy (Fig 3,4). Thus, the conformational
energy of a monomer unit is a double-well potential function for
which the height of the barrier, the distance between the minima,
and the difference in their depth determine the shape of
structural changes in DNA on a given fragment.
\begin{figure}[h]
\centering
\includegraphics[width=0.35\textwidth]{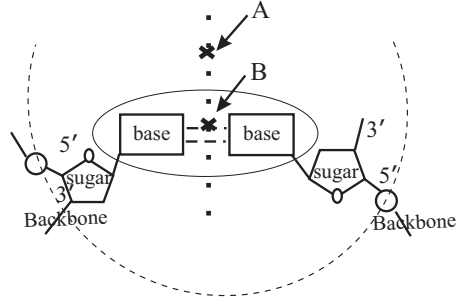}
\caption{Modelling of structural elements of a monomer unit and
their arrangement in pairs relative to the dyad axis.}
\end{figure}
\begin{figure}[h]
\centering
\includegraphics[width=0.43\textwidth]{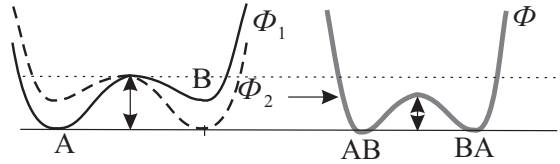}
\caption{ Modelling the potential function of a link
characterizing the mutual arrangement of nucleotides in a pair. A
-is the state corresponding to $C3'-endo$ sugar ring form, B-is
the state corresponding to $C2'-endo$ sugar ring form, BA - is the
state which is characterized with $C3'-endo$ form of sugar ring
for $5'$ and $C2'-endo$ sugar ring form for $3'$ nucleotides of
monomer link, AB - is the state which is characterized with
$C3'-endo$ form of sugar ring for $3'$ and $C2'-endo$ sugar ring
form for $5'$ nucleotides of monomer link}
\end{figure}
The conformational degree of freedom describes relative change in
the positions of the centers of masses of structural elements of
monomer link which is base pair. Figure 3 schematically shows the
structure element of monomer link which consists of pair of
nucleotides (base, five-membered sugar ring, backbone of phosphate
group); position of nucleosides relative to the main axis of the
double helix depicted by $\mathbf{'x'}$ on the dyadic axis in the
plane of the pair for the B and A conformations of the link. The
change in conformations in both nucteosides from the position
corresponding to the A-form to the position corresponding to the
B-form and vice versa is described by the displacement of a pair
of nucleosides within the unit and also leads to a shift in the
center of mass of the nucleosides.

The movement of the link as a whole can also occur without
changing the relative position of structural elements inside the
link and is determined by the displacement of the center of the
link mass with mass $M = 2m_{0} + m_{1} + m_{2}$ and the relative
position of the link masses are described by external components.
The relative mass displacement inside the link is described by the
mass $\mu = m_{1} m_{2} /(m_{1} + m_{2})$ and correspond changes
in the internal conformational component. This approach was
developed and formed as the Volkov-Kosevich model for describing
the vibrations of the structural elements of DNA
\cite{VolkovKosevich} and is successfully used to describe small
vibrations of DNA in various media and complexes, as well as to
describe structural transitions in DNA double helix
\cite{Volkov18}.

According to this approach, each nucleotide is modeled by a
physical pendulum, which can be located in several stable
positions corresponding to different conformations of the sugar
ring. In DNA, the change occurs along a certain conformational
trajectory, the energy of which has two minima corresponding to
two forms of the double helix A and B and is expressed in a change
in the position of nucleosides centers of the mass  in the plane
of the double helix. For the B-form of DNA, the pair is close to
the main axis, and for the A-form it is shifted outward from the
axis, which is clearly seen on X-ray structural images, where the
cavity in the middle is visible in the A-form, while in the B-form
the axis passes through the center of the pairs. Therefore, B-A
changes in the structure are expressed by the joint displacement
of the masses of nucleosides from the center to the periphery as a
whole.

In the case of the TATA box, the shape of the sugar in the
nucleotides of a monomer unit pair is different. Therefore, the
changes in the sugar rings correspond to the rotation of a pair of
nucleosides, which untwisting ensures the arrangement of the pairs
along the axis and the realization of bending while maintaining
the stacking. The potential function will be double-well for
simultaneous BA transition in one chain and AB in another and
differ from the BA transition in both chains by the position and
depth of the wells.

A change in conformation inside a link leads to changes in the
external degrees of freedom and is called internally-induced
deformation. Potential field of DNA deformation forces, consisting
of contributions from elastic deformation, internal conformational
rearrangement, the relationship of these components and the action
of external forces.

The potential energy of the field is determined by the following
types of components: $ U_{el} (R) $ - energy of elastic
deformation (torsion, tension, bending), described by 3-component
deformation $ R $, $ U_{conf} (r) $ - energy of internal
conformational change, is described by the component of the
internal conformational rearrangement of the relative displacement
structural elements within the link, $ r $; $ U _ {\chi} (R, r) $
- potential energy, reflecting the relationship between the
components, $ A (f, R) $ - the work of an external force applied
to one of the elastic components.

\begin{equation} E = U_{el} (R) + U_{conf} (r) + U_{\chi}
(R, r) -A (f, R).
\end{equation}

The general view of the deformation energy of the double helix of
DNA, taking into account the internal rearrangements of the
structure (1), can be used to describe any deformations that are
accompanied by a change in the shape of the sugar ring of
nucleotides. However, we will focus on the possibility of using
this model to describe the deformation at conformationally
bistable fragment as TATA-box is.

\section{Formation of localized DNA strain}

Let us consider the formation of a bend at the region of changing
of the double helix conformation. For the theoretical description
of such deformations, a two-component model of internally induced
deformation can be used, which describes the conformational
transition between two states and the associated bending. The
strain energy density associated with the conformational change in
monomer units can be written as:
\begin{equation}
\varepsilon(u, r) = \frac{C_ {u} u ^ 2} {2} + \frac{C_ {r}r^ 2}
{2} - \gamma_{ur}uF (r) + \frac {\Phi_e (r)} {2},
\end{equation}
where $ u $ is the change in one component of the elastic rod, $ r
$ is conformational change along the chain. $ \Phi (r) $ -double
well conformational function, $ \gamma_{ur} F (r) $ - constant and
potential of the conformation interrelation with double helix
bending. The energy minimum condition determines the shape of the
conformation-induced deformation:
\begin{equation} \frac{\partial\varepsilon(u,
r)} {\partial u} = C_ {u} u- \gamma_ {ur} F(r) = 0;
\end{equation}
\begin{equation} \frac{\partial\varepsilon(u, r)}{\partial r}=\frac{d\Phi_e(r)}{dr} -2 \gamma_ {ur}
\frac{F(r)}{dr} u = 0.
\end{equation}

In the case of a local conformational transition, the change in
the elastic component associated with the conformational component
is also localized and proportional to the potential of the
relationship between the change in the conformational state and
elastic deformation $F(r)$:
\begin{equation} u = \frac{\gamma_{ur}F(r)}{C_ {u}}.
\end{equation}

This solution depends on the shape of the conformational
potential. For a bistable conformational potential, such that the
depth of the wells is the same, the shape of the solutions has the
form of a kink in the conformational component and a bell-shaped
deformation in bending. Kink for the conformational component
corresponds to the fact that the conformational states are
different at the ends of the fragment, and the bell-shaped shape
of the bending component corresponds to the maximum bending in the
center. For the conformational potential with different well
depths, the shape of the solutions has the form of a bell for both
the conformational and bending components, such a deformation
describes the transition of the deformed fragment into A form,
type B-A-B transition. In this way the conformational potential of
the TATA box must have the same well depth.

Another restriction on the coformation-induced bending of the TATA
box is the restriction on the rigidity of the deformed fragment.
It turned out that the internal induced deformation may be more
advantageous than uniform deformation of the same magnitude
without conformational rearrangement. The advantage of the
conformation-induced mechanism is realized in the range of
rigidity of the elastic component
\begin{equation} \gamma_{ur}^2 /
\epsilon <C_ {u} <3 \gamma_ {ur} ^ 2/2 \epsilon.
\end{equation}
According to our previous estimates \cite{KanevskaVolkov2006},
this condition requires almost half the stiffness of the
deformation of the elastic component. Despite the possibility of
internal induced deformation, within the framework of the proposed
model, the reasons for such a strong softening remained unclear,
since the flexural rigidity of the TATA box sequence does not
differ too much from the flexural rigidity of the DNA sequence as
a whole. In the following subsections, we will consider the
mechanism for the formation of bistacility of the conformational
potential, as well as softening of rigidity.

\subsection{Formation of
conformational bistability in the TATA box under the influence of
force}

The mechanism of formation of bistability of conformational states
due to the mobility of the sugar ring was considered in detail for
the threshold DNA lengthening \cite{Volkov18}. In the case of a
TATA box, the external action applied to the ends of the TATA box
is protein loops that intercalate between 1–2 and 7–8 units, offer
unwinding of the fragment and anchoring the sugar shape at the
ends.

According to \cite{Saenger} p.387 during intercalation, structural
changes spread to neighboring pairs to those adjacent to the
intercalator, unwinding along at least three pairs before and
after intercalation. The external forceful action of the iterators
mainly leads to unwinding, and makes the conformational potential
bistable, caused by a change in the forms of sugars (from C3-endo
to C2-endo at one level of severity and from C2-endo to C3-endo at
another severity level). When positioned at a certain distance
between the intercalators, sugar conformation changes using
similar functions, the double helix causes a bend.

A necessary condition for the transition is the possibility of
transition from one form to another under certain boundary
conditions, which are used by intercalators at the ends of the
fragment. Let us formulate 3 conditions for the bending induced by
the conformational transition:

1) bistability of a fragment: bimodality of sugar conformation in
alternating sequences

2) the formation of an unstable conformational state under normal
conditions with the help of external influence. In the case of the
TATA box, the transition between the sugar forms in the strands
caused by intercalators

3) The location of intercalators, as fixers of states, on
distances that are multiples of the length of the cooperativity of
structural changes. The cooperativity in DNA is 3-4 pairs, so the
fragment length is 6-8 bp.

Taking into account torsion and bending, it can be expected that
the unwinding of the fragment under the action of intercalators
also leads to induced bending. Accounting for an external force
can change double-well conformational potential to bistable with
identical wells.

As energy (1) of elastic deformation $ R $, we consider two
coupled elastic components, torsion and bending, $ u, v $,
respectively. Suppose that the force is applied to the torsional
component, and the torsion is associated with the internal
component, the deformation energy density is:
\begin{equation}
\varepsilon(u, v, r) = \frac{C_ {v} v ^ 2} {2}  \frac{C_ {u} u^2}
{2} \gamma_ {vu} vu- \gamma_{vr} vF (r) -\frac{\Phi (r)}{2} -fv,
\end{equation}

where $ C_u, C_v, C_r, $ are the bending, torsional and
conformational stiffness parameters, respectively; $ \gamma_{uv},
\gamma_{vr} $ - parameters of the relationship of torsion with
bending and torsion with the conformational component. The minus
sign in front of the syllable of the relationship of torsion and
conformation reflects a decrease in energy deformation due to
internal mobility. We choose with the sign the syllable twisting
and stretching in the potential generation of deformation, since
an interconnection occurs, which is observed only at large
deformations and does not manifest itself in free DNA.

However, it should be noted that the relationship between the
elastic components can be either positive or negative, depending
on the deformation path \cite{Bustamante}. $ F (r) $ and the
conformational double-well conformational potential, $ \Phi (r) $,
have a similar shape for the state equilibrium with conformational
change equal to 0, $ \Phi (r_{0}) = F(r_{0}) = 0 $. The maximum of
different functions corresponds to the same conformational change
value, $ r = r_{1} $. Another metastable state can be different
and can be transformed into a bistable state by using the
\cite{Volkov18} force. The new state of equilibrium, $ r_ {f} $,
is determined by strength. The state in which the minimum energy
is realized for all degrees of freedom is found by equating to 0
the derivatives with respect to all components of the system:
\begin{equation} \frac {\partial \varepsilon (u, v, r)} {\partial v} =
 C_ {v} v \gamma_ {uv} u- \gamma_ {vr} F (r) -fh =0;
\end{equation}
\begin {equation} \frac {\partial \varepsilon (u, v, r)} {\partial u}=
C_ {u} u \gamma_{uv} v = 0;
\end{equation}
\begin{equation} \frac {\partial \varepsilon (u, v, r)} {\partial r} = \frac {d \Phi
(r)} {dr} -2 \gamma_ {vr} \frac {F (r)} {dr} v = 0.
\end{equation}
By expressing bending strain (4) and substituting it
into expression (3), the expression for expression in the equation
for conformational change (5) obtains the equation of a new
conformational state.
\begin{equation}
\frac {d \Phi (r)} {dr} - \frac {\gamma ^ 2_ {vr}} {\tilde {C} _
{v}} \frac {F ^ 2 (r)} {dr} - 2f \frac {\gamma_ {vr}} {\tilde {C}
_ {v}} \frac {F (r)} {dr} = 0,
\end{equation}
where $\tilde{C}_v = C_v- \gamma ^ 2_ {uv} / C_u $ is effectively
softened torsional stiffness due to the torsion-bending
constraint.

Equation (6) in the absence of an external force satisfies the
condition $ \Phi(r)= \epsilon_{0}{F^{2}(r)}$, therefore,
substituting this expression in (6) we get the expression:
\begin{equation} [(\varepsilon_ {0} - \frac {\gamma ^ 2_ {vr}}
{\tilde {C} _ {v}}) F(r) -fh \frac {\gamma_ {vr}} {\tilde { C} _
{v}}] \frac {F (r)} {dr} = 0;
\end{equation}
The state of the extremum $ r_ {0f} $ determined by the force f
can be found by equating to 0 the first factor in (7)
\begin{equation} F (r_ {0f}) = \frac {\gamma_ {vr} f} {C_ {v} \varepsilon_ {0}
 - \gamma ^ 2_ {vr}} = \frac {\gamma_ {vr} f} {\tilde {C} _ {v}}.
 \end{equation}
And integrating (5) over the conformational component, the energy
macromolecules under the action of sily can be transformed into a
bistable form:
\begin{equation}
\varepsilon(r) =\frac{1}{2}(\varepsilon_
{0}-\frac{\gamma^2_{vr}}{\tilde{C}_{v}})F^2(r)-fh\frac{\gamma_{vr}}{\tilde
{C}_{v}}F(r)+C_{r},
\end{equation}
where $C_ {r} = \frac{\gamma^2_{vr}}{2\tilde{C}_{v}} F(r_{f})$
satisfies the conditions $\varepsilon(r_{f})=0 $ and makes it
possible to rewrite the conformational energy into a bistable
form. Equation (3) can be used to express the twist deformation,
which consists of two parts: caused by the applied force, $ v_f $
and associated with the transformation of the conformational
potential into a bistable state $v_r$:
\begin{equation}
v = v_ {f} + v_ {r}
\end{equation}
\begin{equation} v_f = \frac
{fh} {\tilde {C} _v}, v_r = \frac {\gamma_ {ur} F (r)} {\tilde {C}
_v}.
\end{equation}
\begin{equation}
\tilde{F} = F(r -F r_{0f})
\end{equation}
In the case of the TATA box, the role of force is played by the
intercalation of protein loops, which makes the conformational
potential bistable; moreover, the conformations of the ends of the
TATA box correspond to different states of new potential
\begin{equation}
\Phi_e (r) =\frac{1}{2}\epsilon(F(r)-F (r_{0f}))^2,
\end{equation}
where the lowering of the barrier is described by
the expression:
\begin{equation}
\epsilon =(\varepsilon_0-\frac{\gamma_{vr}^ 2}{\tilde{C}_v}).
\end{equation}

Thus, for a sequence with certain stiffness parameters, there is a
force at which conformational bistability of the double helix
structure occurs. This means that, with a high probability, two
conformational states are realized and the place of the boundary
can be realized in the deformed structure of the DNA chain as a
whole, expressed in deformation and elastic parameters of torsion
and bending.
\begin{center}
\begin{figure*}
\includegraphics[height=35mm]{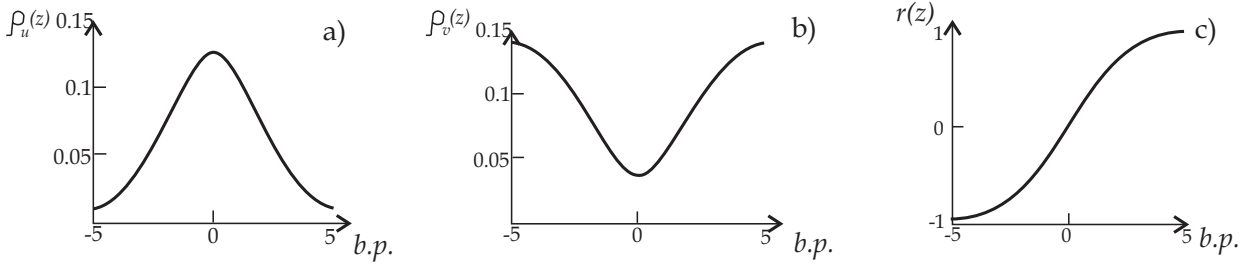}
\caption{a)Twist of each base pair in deformed TATA-box;b)Roll of
each base pair in deformed TATA-box; c)Sugar ring pucker of each
nucleotide in deformed TATA-box}
\end{figure*}
\end{center}
\subsection{Effective softening of bending stiffness in conformation-induced
deformation by twist-bend coupling}

Since the bending of the TATA-box of DNA occurs in the unwound
fragment, and unwinding itself under the action of intercalators
probably leads to bending, in this work we included an additional
elastic component, torsion $ v $, and its relationship with
bending. Since the size of the structural elements is small in
comparison with the chain of the $ h << L $ macromolecule, in
order to obtain analytical expressions describing the
conformational changes and the associated deformation of two
elastic components, we we write down the three-component
deformation model in the continuum approximation.
\begin{equation}
\begin{aligned}
E = \frac{1}{2}\int_{-\infty}^{\infty}\frac{dz}{h}[C_{u}u'^2 + C_{v}v'^2 + gr'^2+\Phi (r)+\\
2\gamma_{ur}u'F (r)+2\gamma_{uv}u'v'],
\end{aligned}
\end{equation}
where $r_{n + 1}-r_{n} =r'$ -conformational change, $u_{n+1}-u_{n}
=u'$, $v_{n + 1} -v_{n}=v'$ elastic deformations bending and
torsion. The Euler-Lagrange equation takes the following form:
\begin{equation}
C_{u} u'' + \gamma_ {uv} v '' + \gamma_{ru} \frac{\partial F
(r)}{\partial r} r '= 0;
\end{equation}
\begin{equation}
C_ {v}v'' +\gamma_{uv} u'' = 0;
\end{equation}
\begin{equation}
gr'' -\frac {1} {2}\frac {\partial\Phi (r)} {\partial r} -
\gamma_{ur} \frac {\partial F (r)} {\partial r} u '= 0.
\end{equation}
From equation (20) $ v'' = -\gamma_{uv}u''/C_{v}$, which,
substituting into Eq. (19) and integrating over z, we obtain an
equation for the deformation shape of the elastic component, which
is suborb to the expression for deformation in two-component
model, but with effectively lower rigidity due to the
interconnection of the urp components $ C_u\rightarrow C_u-\gamma_
{uv}^2/C_{v}$

\begin{equation}
u'= -{\gamma_{ur} F (r)}/{(C_u- \gamma_{uv}^2/C_{v})}.
\end{equation}

When there is no deformation of the molecular chain at the edges
of the deformed fragment, this corresponds to a minimum of the
conformational potential and the following boundary conditions $r
\rightarrow \pm1, ~ r '\rightarrow 0, ~ u' \rightarrow 0, ~ v
'\rightarrow 0, ~ z \rightarrow \infty $. For $ F(r) = 1-r^2 $,
where $ -1 \leq r/r_{0f} \leq 1 $ static solution of equations
(19-21) has the form of a three-component soliton:
\begin{equation}
r(z) = \pm th(\sqrt{Q}z/h);
\end{equation}

\begin{equation}
u (z) = -\frac{\gamma_ {uv}} {(C_ {u} - \gamma_ {uv} ^ 2 / C_ {v})
\sqrt {Q}} th(\sqrt {Q} z / h);
\end{equation}

\begin{equation}
v(z)=\frac{\gamma_{ur}\gamma_{uv}}{(C_{u} -\gamma_{uv}^
2/C_{v})\sqrt{Q}} th(\sqrt{ Q}z /h),
\end{equation}

where $Q = (\epsilon- \gamma_ {ru} ^ 2 / (C_u- \gamma_ {uv} ^ 2 /
C_ {v})) / g $. The form of static deformation is described by a
bell-shaped soliton:
\begin{equation}
\rho_{u}(z) = hu '= - \frac {\gamma_{ur}} {(C_ {u} - \gamma_ {uv}
^ 2 /C_ {v})}ch ^ {- 2} ( \sqrt{Q} z/h).
\end{equation}

\begin{equation}
\rho_ {v} (z) = hv '= \frac {\gamma_ {ur} \gamma_ {uv}} {(C_ {u} -
\gamma_ {uv} ^ 2 / C_ {v})} ch ^ { -2} (\sqrt {Q} z/h).
\end{equation}

According to equation (26), the most deformed part is the central
part of the stressed fragment, corresponding to the maximum
deviation from the conformational minima.

The soliton width is determined from the slope of the tangent at
the center of the conformational transition as $L=2/r'(0)=
2h/\sqrt{Q}$:
\begin{equation}
L = \sqrt {\frac {g}{\epsilon-\gamma_ {ur}^2/(C_u- \gamma_{uv}^2/
C_{v})}}h.
\end{equation}
Due to the dependence of formulas (22-25) on $\sqrt{Q}$, $Q> 0$
and $\epsilon- \gamma_{ru}^2/(C_u-\gamma_{uv}^2/C_{v})>0 $. And we
have the first condition for the parameters of the model.
Substituting solution (23-25) into the energy equation (18) and
integrating over the width of the soliton (27), we obtain the
deformation energy $ E_{ind}\sim(\epsilon- \gamma_{ru}^2/(C_u-
\gamma_{uv}^2/C_{v})) $ that accompanied by a change in the
conformation of monomeric units. Elastic part of the self-induced
bending energy $ E_{el}\sim\frac{1}{2}\gamma_{ru}^2
/(C_u-\gamma_{uv}^2/C_ {v})$
\begin{table*}
\begin{center}
\caption{Range of stiffness in percent for which
conformation-induced deformation is advantageous.The maximum value
of $C_{u}/C_{u_0}$, which intrinsic induced bending has advantage
for. The minimum value of $C_{u}/C_{u_0}$, for which the barrier
between conformational states disappears. Where $C_{u}$ is the
bending stiffness of the deformed fragment, $ C_{u_{0}}$ is the
average bending stiffness.}
\label{tab: 1}
\begin{tabular} {lll}
\hline \noalign {\smallskip} & Max value $C_{u}/C_{u_0}$ & Min value $C_{u}/C_{u_0}$ \\
\noalign {\smallskip} \hline \noalign {\smallskip}
I bend + conformation & 0.54 & 0.36 \\
 II bend + twist + conformation & 0.79 & 0.61 \\
 II + anisotropy & 0.95 & 0.73 \\
\noalign {\smallskip} \hline
\end{tabular}
\end{center}
\end{table*}
\begin{table*}
\begin{center}
\caption{ Parameters of the intrinsic-induced deformation for two
value of the conformational potential barrier}
\label{tab:0}       
\begin{tabular}{lllllllll}
\hline\noalign{\smallskip} ~ &~~ $L, bp$ & ~~~bend, $^o$&~twist,$^o$(II,III)&~~$\frac{C}{C_p}(I)$& ~~$\frac{C}{C_p}(II)$&~~$\frac{C}{C_p}(III)$&~~$\frac{E_el}{L-1}$&~~$\frac{E_ind}{L-1}$\\

\noalign{\smallskip}\hline\noalign{\smallskip}
              ~& ~~~ 4 & ~~~~27 & ~~~~16&~~~~0.6&~~~~0.85&~~~~1 &~~~~0.66&~~~~0.47\\
~$\varepsilon=2.5$ &~~~ 6 & ~~~~52 & ~~~~30&~~~~0.48&~~~~0.73&~~~~0.878&~~~~0.76&~~~~0.18 \\
               &~~~ 8 & ~~~~78 & ~~~~44&~~~~0.45&~~~~0.7 &~~~~0.845&~~~~0.77&~~~~0.1\\
\noalign{\smallskip}\hline
               ~ & ~~~ 4 &~~~~ 34&~~~19&~~~~0.19&~~~~0.72&~~~~0.87&~~~~0.83&~~~~ 0.49\\
 ~$\varepsilon=3$& ~~~ 6&~~~~ 66 & ~~~37&~~~~0.39&~~~~0.64&~~~~0.77&~~~~0.9&~~~~0.17\\
                & ~~~ 8 &~~~~ 91 & ~~~52&~~~~0.372&~~~~0.622&~~~~0.75&~~~~0.9&~~~~0.09 \\
\noalign{\smallskip}\hline
\end{tabular}
\end{center}
\end{table*}
\section{Evaluations and discussions}

To determine the conditions for the implementation of the proposed
deformation scenario, let us estimate the model parameters. To
determine the preference condition for localized deformation
accompanied by conformational rearrangement, $(E_{ind}<E_{el})$.
This is the second condition for the parameters of the model.

The interrelation of the $u;v$ components leads to an effective
softening of the $u$ -component; therefore, the condition for the
advantage of the internal conformational bending mechanism over
uniform elastic bending changes to $\gamma_ {ur} ^ 2 / \epsilon +
\gamma_{uv}^2/C_{v}<C_{u}<3\gamma_{ur}^2/2\epsilon+\gamma_{uv}^2/C_{v}$.

In addition, taking into account the effect of anisotropy in the
bending of the spiral structure leads to an effective decrease in
the rigidity $C_{u} \rightarrow kC_{u}, k <1$. And the stiffness
range that provides the advantage of the internally induced
bending mechanism expands: $(\gamma_ {ur} ^ 2 / \epsilon + \gamma_
{uv} ^ 2 / C_ {v}) / k <C_ {u} <(3 \gamma_ {ur} ^ 2/2 \epsilon +
\gamma_{uv}^2/C_{v})/k$. According to our estimate, for the
TATA-box sequence, $ k = 0.83 $. This value matches the range
anisotropy effect described in \cite{Volgodskii_book}.

Thus, the relationship between the elastic components and the
anisotropy of the helical structure opens up opportunities for the
emerging conformational transformations in elastic components. To
estimate the range of advantages of the internal induced
mechanism, we used the following values of the model parameters.
Flexural stiffness $C_{u} =94~kcal/mol$, torsional stiffness
$C_{v} = 75 ~ kcal/mol$, relationship between torsion and bending,
$\gamma_{uv} = 0.5 \sqrt {C_ {u} C_ {v}} = 42 ~ kcal/mol$.

The parameters of conformational rigidity, the barrier between
conformational states, and the relationship between conformation
and bending were evaluated in a previous study [5]. They are
respectively: $g = 20 ~ kcal / mol, ~ \epsilon = 3 ~ kcal / mol, ~
\gamma_ {ur} = 0.6 \sqrt {\epsilon C_u} =10 ~ kcal/mol$.

For the given parameter values, we calculate the range in which
the conformation-induced mechanism has the advantage of Table 1.
Model I includes bending and conformational change, model II adds
the relationship of torsion and bending, model III also takes into
account the bending anisotropy. The ratio of the fragment rigidity
to the average bending rigidity of the DNA double helix
$C_{u}/C_{u_0}$ shows the degree of deviation of the bending
rigidity parameter of a DNA fragment for localization of
conformation-induced deformation in it.

In a simple two-component model, the deformed fragment must
undergo significant changes, up to local destruction. The
assumption that the elastic components are interconnected leads to
a more probable internal induced deformation. Finally, the
inclusion of the influence of anisotropy makes it probable that
internal induced deformation appears in a fragment with $ 5\% $
softened rigidity.

More detailed results of applying the deformation models are
presented in Table 2. The effect of reducing the barrier of the
conformational potential under the action of the untwisting force
of intercalators obtained in Section 3.1, formula 19 is also
considered. In this case, the barrier is $\epsilon = 2.5 ~
kcal/mol$. bending, unwinding, softening of the fragment
$C_{u}/C_{u_0}$ for three types of models, as well as the energy
per base pair at elastic uniform deformation with the same
softening of the fragment stiffness and the average energy per
pair at the conformationally induced localized bending of the DNA
fragment.

All values are given for localized deformations on 4,6,8 base
pairs. As can be seen from Table 2 for the third model, which
takes into account both the internal component and its
relationship with the external one and the relationship between
internal components, the anisotropy of bending and the effect of a
decrease in the barrier between conformational states during the
formation of bistability under the action of the untwisting force
intercalators, without softening the stiffness for 4 pairs and
with about $12 \%$ reduced stiffness, localized bending can form
according to the conformation-induced mechanism.

It is interesting to note that for a localized deformation with a
length of 6 pairs, the bending magnitude is $52^o$ and untwisting
by $27^0$ provided that the flexural stiffness decreases by
$12.2\%$, which is realized due to the neutralization of charges
by the protein complex. Thus, the general conformation-induced
deformation of the central fragment of the TATA box into $ 95 ^ 0
$ untwisting and $85^0$ bending is realized.

In the paper the model of deformation of conformationally bistable
DNA fragment was formulated. The model allows to describe
consistent mechanism of deformation of the DNA TATA box  during
binding with protein complex.  The maximum deformation in the
center and its stability sufficient for proteins to recognize it
is argued. The main elements of the  model of intrinsic-induced
DNA deformation are bistability of the conformational potential of
the DNA fragment, the coupling of internal conformational changes
with a change in parameters determining twist and bending
deformation, the coupling between torsion and bending and effect
of bending anisotropy of double helix DNA. These elements form the
physical basis of the unique deformation of the TATA-box of DNA
and make it possible to explain and predict the mechanisms of
controlled DNA compaction.

{}
\end{document}